# Modeling Fractal Structure of Systems of Cities Using Spatial Correlation Function

Yanguang Chen[1], Shiguo Jiang[2]

(1. Department of Geography, Peking University, Beijing 100871, PRC. E-mail: chenyg@pku.edu.cn;

2. Department of Geography, The Ohio State University, USA. Email: jiang.152@osu.edu.)

**Abstract**: This paper proposes a new method to analyze the spatial structure of urban systems using ideas from fractals. Regarding a system of cities as a set of "particles" distributed randomly on a triangular lattice, we construct a spatial correlation function of cities. Suppose that the spatial correlation follows the power law. It can be proved that the correlation exponent is the second order generalized dimension. The spatial correlation model is applied to the system of cities in China. The results show that the Chinese urban system can be described by the correlation dimension ranging from 1.3 to 1.6. The fractality of self-organized network of cities in both the conventional geographic space and the "time" space is revealed with the empirical evidence. The spatial correlation analysis is significant in that it is applicable to both large and small sizes of samples and can be used to link different fractal dimensions in urban study, including box dimension and radial dimension.

## 1 Introduction

The evolution of cities as systems and systems of cities bears some similarity. In theory, a system of cites follows the same spatial scaling laws with a city as a system. The great majority of fractal models and methods for urban form and structure are in fact available for systems of cities. The theory of fractal cities was systemized by Batty and Longley (1994) and Frankhauser (1994) more than ten years ago. From then on, a number of advances have been made (e.g. Batty, 2008; Benguigui *et al*, 2000; Benguigui *et al*, 2006; De Keersmaecker *et al*, 2003; Thomas, *et al*, 2008). The fractal concepts on cities as systems are well developed, and the results provide the theoretical foundation for the study of fractal theories on systems of cities.

However, a scale difference comes between systems of cities and cities as systems clearly.

Systems of cities are defined at the macroscale associated with interurban geography, while cities as systems are defined at the microscale associated with intraurban geography. There seems to be an explanatory breakup between the macroscale and the microscale. On the one hand, we cannot use the spatial interaction of intraurban elements (e.g. housings) to explain interurban dynamics and the spatial order of self-organized network of cities. On the other, we cannot use the rule based on the large scale, say, national scale, to explain the city growth and urban form. Therefore, it is necessary to make fractal studies on systems of cities. It will help us understand urban evolution by comparing the rules of these two scales of systems.

An important point in urban studies of large scale is the rank-size rule. The Zipf's law in essence is a kind of fractal model (Chen and Zhou, 2003; Frankhauser, 1991). The fractal dimension of the rank-size distribution of cities is theoretically the ratio of the dimension of urban network and that of regional population distribution; it is also the ratio of the dimension of urban form and that of population distribution of the largest city in an urban system (Chen and Zhou, 2006). This discovery is revealing for us to bright to light the relationship between intraurban fractal models and interurban fractal models. Although there are not as many fractal studies on systems of cities as those on fractal cities, the fractal studies of central place (e.g. Arlinghaus, 1985; Arlinghaus and Arlinghaus, 1989; Chen and Zhou, 2006), the rank-size distribution (e.g. Chen and Zhou, 2004; Frankhauser, 1991; Wong and Fotheringham, 1990) and transportation network (e.g. Benguigui and Daoud, 1991; Chen and Liu, 1999; Dendrinos and El Naschie, 1994; Lu and Tang, 2004), etc., make a good outset for further research.

The well-known difficulty in the fractal study of systems of cities is the sample size problem. For a small regional scale, we have small number of cities and small spatial sample size. Under such circumstances, we cannot use the box-counting method to compute the fractal dimension. So it is hard to reveal the fractal structure of systems of cities. The spatial correlation analysis is a feasible way out of this difficulty. In this paper, based on the spatial correlation function, we develop a method to build the fractal model for systems of cities. The spatial correlation model is important for urban studies in that it possesses a potential to link several kinds of fractal models. First, the number-scale scaling and the area-radius scaling relation commonly used in fractal urban studies are actually special cases of correlation function. Second, the correlation dimension and the box dimension belong to the same dimension spectrum. The box dimension is the zero-order dimension and the correlation dimension the second order dimension. In this sense, the correlation function not only is helpful in the fractal study of urban systems with small sample size, but also has the potential to build a logical relation between different fractal models.

The correlation function is an effective mathematical tool to describe the fractal structure of the chaotic attractor in pseudo phase space. It is simple yet natural to extend the correlation function analysis from the phase space to the real geographical space. The difficulty lies in how to prove the validity of this extension both in theory and practice. The contribution of this paper can be summarized as follows. First, we prove that the spatial correlation exponent is just the second order information dimension based on the box-counting method. Second, we give a complete example of making use of the spatial correlation model. Third, we provide a systematic analysis on the spatio-temporal evolution of systems of cities using the fractal dimension and the scaling range. Fourth, we demonstrate the relationship between the spatial correlation dimension and the radial dimension of systems of cities.

The remainder of this paper is structured as follows. Next section presents a spatial correlation model. The scaling exponent of the model is proved to be just the spatial correlation dimension of urban systems. Section 3 provides an empirical case to illuminate the spatial correlation model by applying it to the principal cities in China. The geographical meaning of the scaling range and the correlation dimension are illustrated with three measures of urban networks. Section 4 expands the discussion about some related questions, especially to demonstrate the relationship between the spatial correlation dimension and the radial dimension. Finally, this paper is concluded with a summarization of the main points of the modeling approach to fractal systems of cities.

## 2 Theoretical and model

### 2.1 Triangular lattice model of urban system

The way we compute the fractal dimension based on the mass correlation function of urban system in geographical space is similar to that of computing the correlation dimension of strange attractors in phase space. The latter was developed by Grassberger and Procaccia (1983a, 1983b). However, the basis of building model is different. Considering urban systems to be networks of cities and towns, we start here with the triangular lattice model of city distribution, which is connected with the fractal structure of central place (Chen and Zhou, 2006). As Batty and Longley (1994, page 48) point out, "The simplest geometric form of a system of cities is based on an entirely regular grid of basic settlement types – neighborhoods or villages say – which are systematically aggregated into all encompassing regions at successive levels up the hierarchy." According to this idea, the geographical surface can be thought of as a sort of triangular lattice, while cities are particles distributed randomly on the lattice.

The triangular lattice analogy is made for two reasons: (1) it is consistent with the central place model of Christaller (1966); and (2) it is in agreement with the study of spatial complexity. Some researchers such as Bura *et al* (1996), Sanders *et al* (1997) and Sanders (2006) take triangular lattice as a spatial support in the study of settlement system and urbanism by using the multiagent system (MAS). In modern science, the complexity is thought of as the order in a random background. Correspondingly, in our assumption of the triangular lattice, the urban system is treated as the random distribution in the orderly background (Figure 1). The order in the random background and the random distribution in the orderly background are in fact dual questions.

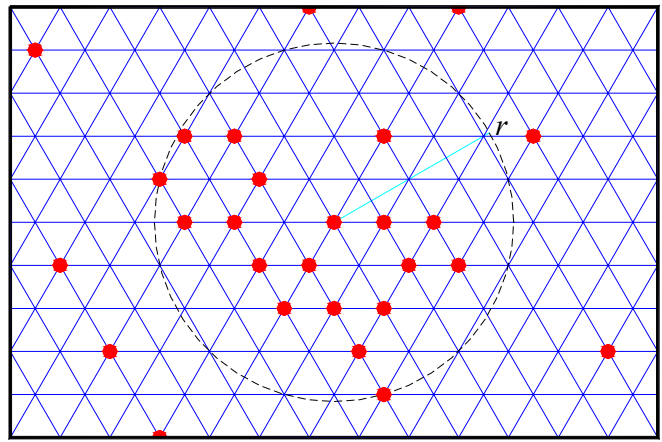

**Figure 1. The triangular lattice pattern of the spatial distribution of an urban system**
**Note**: The background of the model is a triangular lattice, which is consistent to the central place network. The real cities can be thought of as random particles distributed on the standard lattice nodes. This kind city distribution no longer has the rigid symmetry. The number of coordinate cities around a city does not equal 6.

The notion of triangular lattice model of urban systems can be developed as follows. First, take the ideal earth surface as a triangular lattice. Original settlements are well-distributed on the lattice nodes, forming a triangular lattice. Secondly, there exist some unpredictable random factors, namely *chance factors*, which cannot be expressed by models. Owing to the effect of chance, some original settlements evolved into cities or towns earlier than others. The immediate settlements around the cities are restrained from growing into cities because of "shade effect" (Chen and Zhou, 2006; Evans, 1985). To outward seeming, these cities are distributed randomly on the lattice nodes. The first aspect indicates a kind of determined geographical spatial system, while the second one indicates the random activity of city evolution in the deterministic system.

We start our work of modeling spatial correlation of cities from the first law of geography proposed by Tobler (1970, P236): "Everything is related to everything else, but near things are more related than distant things."(See also Tobler, 2004, P304) We can derive two hypotheses from this law. One is the *spatial correlation effect*: every city is related to every city else. The other is the *distance-decay effect*: near cities are more related than distant cities. The spatial correlation effect is the basis for our modeling, while the distance-decay effect is the prerequisite for mass correlation function. There are two basic modes of the spatial action in urban evolution. One is *action at a distance*, suggesting an inverse power-law relation; the other is *locality*, suggesting a negative exponential distribution (Chen, 2008). If the above hypotheses are true, the correlation function of cities will follow some power laws, and the correlation exponent is the fractal dimension or the function of fractal dimension.

In the triangular lattice model shown in Figure 1, every city can be counted as a particle. The problem is to find the regularity in the spatial distribution patterns of these "particles". For purpose of this, it is necessary to make some mathematical models. By using the concepts of auto-covariance and standardized data (e.g. Dendrinos and El Naschie, 1994; Vicsek, 1989), we can define a density-density correlation function such as

$$R(\boldsymbol{r}) = \frac{1}{S} \sum_{\boldsymbol{x}} \rho(\boldsymbol{x})\rho(\boldsymbol{x}+\boldsymbol{r}), \tag{1}$$

where $R(\boldsymbol{r})$ denotes the density correlation, a generalization of the covariance function, $\rho(\boldsymbol{x})$ refers to city density, $S$ represents the area of a cellular unit occupied by a particle, $\boldsymbol{x}$ is the location of a certain particle (defined by the radius vector), and $\boldsymbol{r}$ denotes the distance to $\boldsymbol{x}$. According to equation (1), if there is a city at $\boldsymbol{x}$, the probability of finding another city at distance $\boldsymbol{r}$ from $\boldsymbol{x}$ is $R(\boldsymbol{r})$. The key problem is how to determine the relationship between correlation function $R(\boldsymbol{r})$ and distance $\boldsymbol{r}$.

Suppose that the correlation function satisfies the scaling relation (Feder, 1988; Vicsek, 1989)

$$R(\lambda \boldsymbol{r}) \propto \lambda^{-\alpha} R(\boldsymbol{r}). \tag{2}$$

Thus the system has the property of scaling invariance, and its structure is self-similar in certain spatial-temporal condition. In equation (2), $\lambda$ is the *scale ratio*, and $\alpha$ is the *scaling exponent* ($\alpha>0$). If and only if the density correlation function takes a form of inverse power law such as

$$R(\boldsymbol{r}) \propto \boldsymbol{r}^{-\alpha}, \tag{3}$$

will the functional equation (2) come into existence. In this case, the scaling exponent is expected to be $\alpha=d-D$, where $d=2$ is a Euclidean dimension of geographical space, and $D$, the correlation dimension of cities. Equation (3) is a spatial scaling law. In fact, it has been shown that the correlation density follows the power-law distribution or the exponential distribution, sometimes other distributions. What kind of distribution will it be as to a system of cities? This cannot be proved with *apriori* knowledge. It depends on the real empirical results. If an urban system evolves into a self-organized critical state, the spatial correlation follows a power law. Otherwise, it follows an exponential law (Liu and Liu, 1992). More discussion can be seen from Batty and Kim (1992) for the comparison of exponential laws vs. power laws.

The density correlation function is useful in characterize urban form, the DLA models, etc., but it seems not to be a convenient modeling approach to the strange attractors in a two- or three-dimensional pseudo phase space (Williams, 1997). In order that the correlation function can be applied to the two-dimension pseudo phase space, the density correlation should be transformed into the mass correlation. For simplicity, we assume that the fractal object is of isotropy, i.e., the correlation of particles is independent of direction. The density correlation function only depends on the **scalar** $r$, equation (2) reduces to a homogeneity relation, and $R(\boldsymbol{r})=R(r)$. Defining a cumulative distribution of city density, we have a mass function

$$A(x+r) = 2\pi \int_0^r \xi \rho(x+\xi) \mathrm{d}\xi, \tag{4}$$

in which $\xi$ is a spatial displacement ($0\leq\xi\leq r$, $r>0$). For a certain point $x$ (i.e. $r=0$), the density approaches to infinity. $A(x)$ can be endowed with a value by taking a unit area for it so that the point is of mathematical meaning. Therefore, the density function has no difference from the mass function, i.e., $\rho(x)\equiv A(x)$. In this instance, the density notation $\rho(x)$ can be replaced by the mass symbol $A(x)$, and equation (1) can be changed into an integral form

$$C(r) = 2\pi \int_0^r \xi R(\xi) d\xi = \frac{2\pi}{S} \sum_x \rho(x) \int_0^r \xi \rho(x+\xi) d\xi = \frac{1}{S} \sum_x A(x) A(x+r). \qquad (5)$$

where $C(r)$ is called **correlation integral** or **correlation sum** (Williams, 1997). Thus the density correlation, a decreasing function, is transformed into the mass correlation, an increasing function.

Generally speaking, in equation (5), $x$ is location variable, $r$ is a scale parameter. However, for the scaling analysis, the variable $x$ can be fixed and $r$ acts as a variable. The attribute of $r$ depends on the analytical framework. We can change $r$, for example, by multiplying it with a numerical value $\lambda$. If the structure of urban system is simple, i.e., it has certain characteristic length, the mass-mass correlation function, equation (5), will change with the parameters. Thus we can find a most proper value for $r$ as a constant (no scaling invariance). If the structure of urban system is of singularity, equation (5) implies the scale-free characteristics. Therefore, for a certain system, regarding all locations as known numbers, and the parameter $r$ as a variable (instead of constant), we can get mathematical relation between the correlation function $C(r)$ and the displacement scale $r$, based on which we can compute the correlation dimension.

For convenience, the triangular lattice state can be divided into two types: if there exists a particle, the state is 1; otherwise, the state is 0. Then we can use the dummy variable to describe the spatial distribution of a system of cities.

## 2.2 Spatial correlation sum function

In the following part we will derive the correlation sum function in the geo-spatial view. For a homogeneous region with $N$ cities, if there is a city at $x$, the state of $x$ is considered to be 1 in terms of our assumption, that is, $A(x)\equiv 1$. As to the field within a radius of $r$ from $x$, there are two possible states: (1) cities occur in the scope, the state is represented by 1; (2) no city in the scope, the state is 0. Given homogenous distribution of the $N$ points, the area $S$ can be expressed by the number of cities $N$ in the topology sense ($S \sim N^2$). In fact, taking a city $i$ at $x$ as the center, we can draw a circle $\mathbf{C}_i$ with a radius of $r$ to represent the field, and then examine the probability that city $j$ is compassed by circle $\mathbf{C}_i$. Filling the circle forms a disk, which indicates a small box on the digital map. For city $i$, it is surely in the circle, and the probability $P_i(x)=A(x)/N=1/N$. As to city $j$, there are two kinds of possibilities: (1) it is in the circle, the probability $P_j(x)=1/N$; (2) it is not in the circle, the probability $P_j(x)=0/N$. So the cumulative probability of finding city $j$ within the circle can be represented in the Heaviside function form such as $A(x+r)/N=\sum H(r-d_{ij})/N$, where $d_{ij}$ is the distance between city $i$ and city $j$, and $H(\cdot)$ is the Heaviside function. Thus, for the standardized data, equation (5) can be replaced by

$$C(r) = \frac{1}{N^2} \sum_{i=1}^{N} \sum_{j=1}^{N} H(r - d_{ij}), \tag{6}$$

where $r$ is the length of the yardstick, corresponding to the displacement parameter, $C(r)$ is the spatial correlation sum. The Heaviside function has the following property

$$H(r - d_{ij}) = \begin{cases} 1, & \text{when } d_{ij} \leq r; \\ 0, & \text{when } d_{ij} > r. \end{cases} \tag{7}$$

Apparently, $H(\cdot)$ acts as a kind of gatekeeper or admissions director. If the density correlation function follows the inverse power law in the form of equation (3), i.e. $R(r) \propto r^{-(d-D)}$, we have $C(r) \propto r^{-(d-D)+2}$ according as equation (4). When $d=2$, we have

$$C(r) \propto r^D, \tag{8}$$

where $D$ is the scaling exponent, i.e. the correlation exponent (Williams, 1997). It can be proved that $D$ is just the correlation dimension, and the value of which varies from 0 to 2.

## 2.3 The dimension implications of the scaling exponent

The correlation dimension implications of the aforementioned scaling exponent can be expounded through the generalized dimension. Based on the Renyi information entropy (Rényi, 1970), the generalized correlation dimension $D_q$ can be defined as follows (Feder, 1988; Vicsek, 1989)

$$D_q = \frac{1}{q-1} \lim_{\varepsilon \to 0} \frac{\ln \sum_{i=1}^{N(\varepsilon)} P_i^q}{\ln \varepsilon}, \tag{9}$$

where $q$ refers to the order of moment ($q = -\infty, \cdots, -2, -1, 0, 1, 2, \cdots, \infty$), $\varepsilon$ to the dimensionless box size, and $P_i$ to the count in each box divided by the total count and thus the summation is over the $N(\varepsilon)$ boxes. Equation (9) gives a multifractal spectrum (Mandelbrot, 1999), and it is a reference formula for the following analogy.

In order to prove that the scaling exponent $D$ in equation (8) represents the correlation dimension, we conduct the following spatial analysis. Using any of the cities, say city $i$, as the center to draw a circle of $C_i$ with a radius of $r$, we then examine the probability to find cities in the circle (Figure 2). Suppose that $C_i$ compasses $N_k(i)$ cities, including the center city $i$, and $k$ can be regarded as the serial number of all the cities compassed by $C_i$. Then the cumulative probability of finding cities at a randomly chosen site within the given circle is

$$C_i(r) = \frac{1}{N}\sum_j H(r-d_{ij}) = \frac{N_k(i)}{N} = P_k(i), \qquad (10)$$

In terms of equation (6), the spatial correlation function can be expressed in the form

$$C(r) = \frac{1}{N}\sum_i C_i(r) = \frac{1}{N}\sum_i P_k(i), \qquad (11)$$

If $C_i$ compasses city $j$, $C_j$ will compass city $i$ (Figure 2a). Now cities $i$ and $j$ share the same serial number (Figure 2b). In this instance, all the cities compassed by the circle of radius $r$ can be compassed by a small box with the same scale $r$. The box is numbered by $k$, and the number of small box $N_r \leq N$. Then the spatial correlation function equation (11) can be revised as

$$C(r) = \frac{1}{N}\sum_i \frac{N_k(i)}{N} = \frac{1}{N}\sum_k N_k(i)\frac{N_k(i)}{N} = \sum_k P_k^2, \qquad (12)$$

where $P_k = N_k(i)/N$ is the probability of city-city correlation depending on scale $r$. Then equation (8) can be rewritten as

$$D_s = \lim_{r \to 0} \frac{\ln \sum_k^{N_r} P_k^2}{\ln r}. \qquad (13)$$

A subscript, $s$, is attached to fractal dimension $D$ to denote spatial correlation. Comparing equation (13) with equation (9), we can find that the scaling exponent $D_s$ is just a kind of correlation dimension, which can be termed as *spatial correlation dimension*.

The value of spatial correlation dimension comes between 0 and 2. In fact, for an isolated city in the region, we have $P_k = P = 1/N_r = 1$. The length of the smallest yardmeasure is close to 0 infinitely. According as equation (13), we have

$$D_s = \lim_{r \to 0} \frac{\ln \sum_k^{N_r}(1/N_r)^2}{\ln r} = \lim_{r \to 0} \frac{\ln[(1/N_r)^2]}{\ln r} = 0; \qquad (14)$$

If cities distribute evenly in triangular array in a 2-dimension plane, still we will have $P_k=P=1/N_r$, and the length of the smallest yardstick is $r=1/\sqrt{N_r}$. According as equation (13), we have

$$D_s = \frac{\ln \sum_{k}^{N_r}(1/N_r)^2}{\ln r} = \frac{\ln[N_r(1/N_r)^2]}{\ln(1/\sqrt{N_r})} = 2. \qquad (15)$$

Apparently an even distribution system suggests a trivial structure of Euclidean geometry. The dimension of the cities distributed equably on the triangular lattice can be regarded as 2. It has been proved that the fractal dimension of the standard central place pattern is just $D=2$ (Chen and Zhou, 2006). However, the spatial distribution of real cities has a dimension varying from 0 to 2.

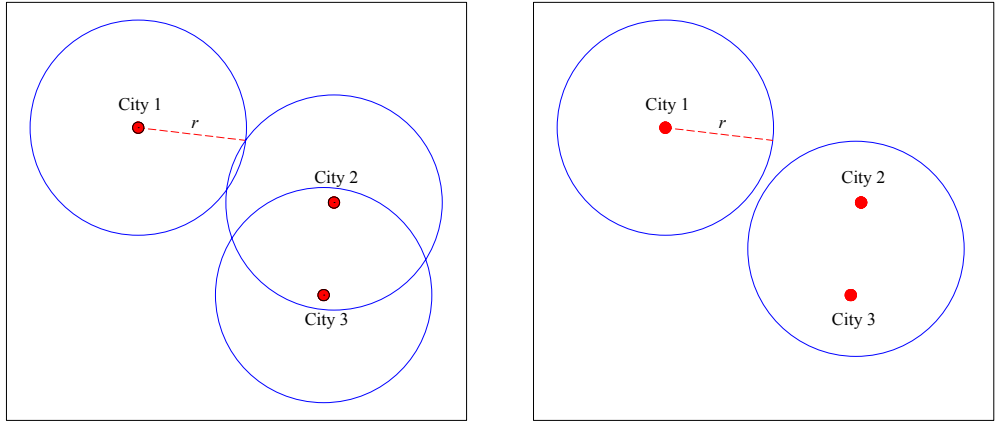

a. Spatial correlation of cities  b. Coverage of small box

**Figure 2. Demonstration on the spatial correlation of cities and the coverage of small boxes**

**Note**: In Figure 2a, the serial number of the three cities are $i, j$=1, 2, 3. The number of cities is $N$=3. Using each city as the center, we can draw a circle $C_i$ with a radius $r$ ($C_i$ also represents a box). In the given scale $r$, $C_1$ compasses only city 1, while $C_2$ compasses city 2 and city 3 and $C_3$ compasses city 3 and city 2. Therefore, if we cover the cities using small boxes with the scale of $r$, we need only two boxes, as shown in Figure 2b. City 2 and city 3 will share the same small box. The serial number of boxes can be defined as $k$=1, 2. The number of small boxes is $N_r$=2<$N$=3 (refer to Appendix 1).

## 3 Empirical analysis: a case of China

### 3.1 The size threshold of cities and distance measurement

Compared to the study of fractal cities as systems, the research on systems of cities poses a much harder challenge to us. Firstly, it is difficult to define the limits of space and the size of an urban system owing to the self-similarity of both urban form and the rank-size distribution of cities (Batty and Longley, 1994; Chen and Zhou, 2003; Frankhauser, 1994). Secondly, geographical laws are the statistical laws which always appear at the macro level of systems. Geographical regularity of urban systems can be effectively revealed by large samples with a great many cities, as indicated by the work of Zanette and Manrubia (1997). Thirdly, geography seems to be a science of process instead of a science of being/existence. It always takes a long time of evolvement for geographical features to reach a state of regularity. The geographical laws should be understood in both space and time context. China is an available place for us to find statistical spatial regularity in that it has a large sample of cities and a long history. Our study area covers the whole continent part of China, including more than 660 cities (Figure 3).

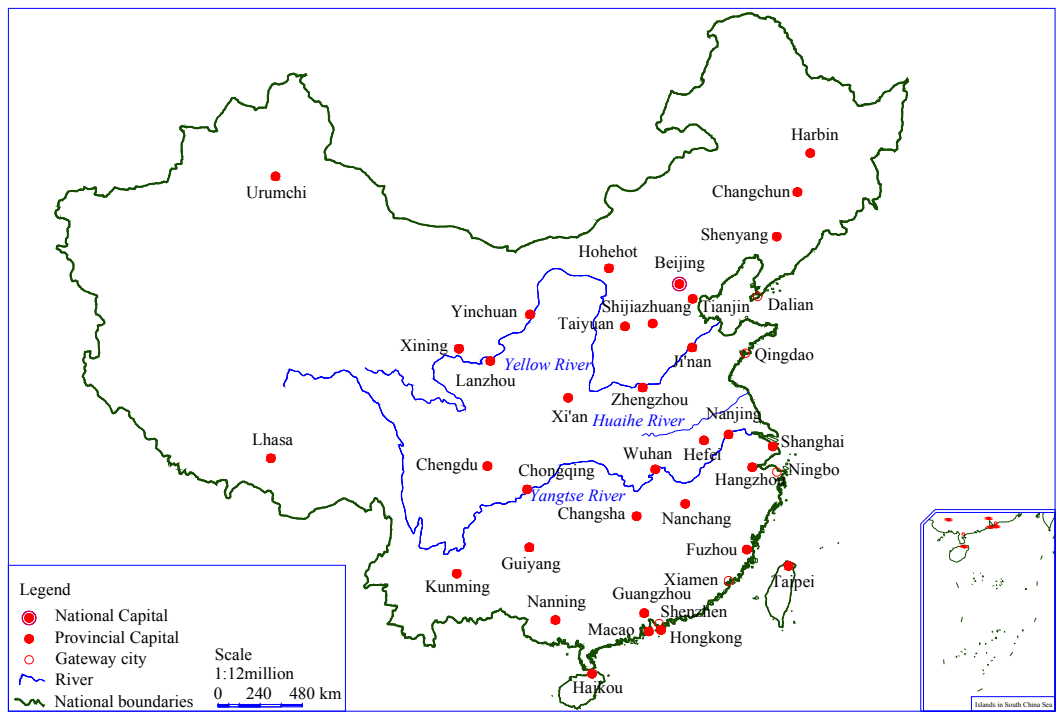

**Figure 3. The spatial distribution of the main cities in China**

**Note**: All the gateway cities except Shenzhen are important seaports in China. They are big cities similar to provincial capital cities. Their city function is as important as that of provincial cities. Although Shenzhen is not a seaport, it is a gateway connecting the mainland and Hongkong.

The lower limit of city size should be defined for our analysis. To estimate the lower size limit is analogous to determining a level of "resolution". For an urban system obeying the Zipf's law, the city-size distribution is of scale invariance, and theoretically there is no distinct limit between urban and rural settlements. Zipf's law indicates fractal structure (Chen and Zhou, 2004; Frankhauser, 1990; Nicolis, *et al*, 1989; Wong and Fotheringham, 1990), while a fractal implies that the scale can be very small. However, in practice, we can discriminate city from non-urban settlements by the scaling range of the double logarithmic plot demonstrating the rank-size distribution of cities, because the self-similarity always break down when the scale is too large or too small. Actually, it is demonstrated that the rank-size distribution of rural settlements follows the exponential law instead of the power law (Grossman and Sonis, 1989; Sonis and Grossman, 1984).

The lower limit of the scaling range of the power-law distribution of cities is the critical size threshold, the tail below which can be regarded as non-urban settlements. Among all the cities (more than 660) in China, about 510 cities appear above the critical size threshold. However, it is difficult to demonstrate the analytic process in pictures if we consider all the 510 or so cities in computation. For simplicity, we consider the most outstanding cities according as both the city function and size. Thus the number of cities can be reduced to $N$=39, including the 34 cities as provincial capitals and the 5 cities as the most important gateways (Figure 3). So we will treat with a 39×39 distance matrix.

Note that the number of spatial correlation is a concept easy to be misunderstood. For a system of $N$ cities, we have $N \times N$ square matrix. Clearly the number of correlation is $(1/2)N(N+1)$. Excepting the $N$ diagonal elements indicative of the auto-correlation distance, we get a number of correlation of $(1/2)N(N-1)$. However, it is hard to understand the correlation number of $(1/2)N(N-1)$ in physical meaning. For the spatial correlation of any two cities, say city $i$ and city $j$, the number of correlation is not 1 but 4. Here we have the cross-correlation of $i$ and $j$ (based on $i$), $j$ and $i$ (based on $j$), auto-correlation of $i$ and $i$ (based on $i$), and $j$ and $j$ (based on $j$). Therefore, in the correlation probability function, the denominator is $N^2$ instead of $(1/2)N(N-1)$. Otherwise, we cannot implement the related mathematical transformation for the spatial correlation model.

Distance is one of the basic measures in the urban spatial analysis. We have two kinds of distances: one is in a straight line, the other in a curve line. The former is equivalent to the *crow distance*, while the latter may be analogous to the *cow distance* (Kaye, 1989, P178). In terms of our triangular lattice model, the straight-line distance between cities $i$ and $j$ can be defined in the system of barycentric coordinates first introduced by Moebius (1827). The distance formula is

$$d_{ij} = \frac{1}{\sqrt{2}} \sqrt{(x_i - x_j)^2 + (y_i - y_j)^2 + (z_i - z_j)^2}$$
$$= \sqrt{(x_i - x_j)^2 + (y_i - y_j)^2 + (x_i - x_j)(y_i - y_j)}, \qquad (16)$$

where $d_{ij}$ denotes a beeline, $x$, $y$ and $z$ are the coordinates of cities. Equation (16) is equivalent to the definition of the Euclidean distance based on the rectangular coordinates. The distance can be computed easily by using digital maps in GIS. Obviously, if we use the beeline distance, the geographical implication of the probability $C_i(r)$ is the ratio of the number of cities falling into the circle of radius $r$ centered at city $i$ to the number of all the cities $N$. According as equation (11), the correlation function $C(r)$ is the sum of $C_i(r)$ divided by $N$.

As to the city distribution in the real world, however, it is very complex. The geographical implication of beeline distance is more explicit to demonstrate on a map, while the transportation distance has more practical sense. Therefore, for an empirical analysis, we extend from the Euclidean distance to the real transportation distance, including the distance by highway or railway between any two cities. In this case, the above circle covering cities will have topological transmogrification.

As mentioned above, the correlation exponent based on beeline distance is just the correlation dimension in two-dimensional embedding space. However, how to interpret the correlation exponent based on transportation distance? We need to refer to the concept of multidimensional scaling of coordinate and the remapping method. In principle, if only we have a distance matrix of a certain kind of nodes or location, measured by Euclidean distance, journey time, travel cost or transportation distance, we can remap the spatial distribution. As Haggett *et al* (1977, P326) pointed out: "Geographers are frequently concerned with discovering map structure from a matrix of interpoint distances." Although there is no precise solution to this mapping problem, there are ways of making a 'best estimate' of the cities' location (Haggett, 2001). As the railway distance and highway distance are generally proportional to travel time, we can use the method of multidimensional scaling to transform a conventional transportation map into a new map format. In the new map, the beeline distance among cities equal to certain transportation distance, and we can use the same method as the above-shown one to prove the correlation dimension in the "time" space (See Haggett *et al*, 1977, for the concept of time space).

Here we do not need to carry out this kind of map transformation in that we have a better method to compute the correlation dimension based on transportation distance. However, such ideas of multidimensional concepts and spatio-temporal map transformation provide us with a theoretical foundation to interpret the spatial correlation: the correlation dimension based on beeline distance is a kind of fractal dimension in conventional geographic space, while the correlation dimension based on transportation distance is in essence a kind of fractal dimension based on a time-space map.

The following is an empirical analysis of China's cities by using our method. The distance data by both railway and highway can be gotten from the Chinese transportation map. Owing to the absence of railway or highway data between some cities, our computation based on the railway network will deal with only 31 cities (a $31 \times 31$ distance matrix), and the computation based on the highway network involves 34 cities (a $34 \times 34$ distance matrix). In fact, we are in a dilemma. On the one hand, the number of cities should be kept constant in order to make comparable the fractal parameters based on different distances; on the other, the elements of the urban system should be as many as possible in order to benefit the future study related to this subject. As a compromise, the computation based on different distances will involve different numbers of cities.

### 3.2 Computation method and results

It is easy for us to deal with the spatial correlation function based on the Euclidean distance by drawing circles with the aid of GIS (see Longley *et al*, 2001). However, this method seems to be unsuited for the distance by highway or railway. In order to avoid difficulty involving curve length (transport mileage), we use the yardstick to measure the number of cities. Thus, the work of drawing circles on digital map (representing concrete space) is replaced by counting in table (representing abstract space). Here we investigate the regularity in the spatial distribution of cities using the Euclidean distance. For a system of 39 cites ($N$=39), the distances between any two cities form a square matrix with dimension 39×39, which has $N^2$=1521 elements. The entries are the distance $d_{ij}$. For simplicity, let $N(r)=C(r)N^2$, i.e., we substitute the accumulative number $N(r)$ for the accumulative density $C(r)$; Therefore equation (6) changes to

$$N(r) = \sum_{i=1}^{N} \sum_{j=1}^{N} H(r - d_{ij}), \qquad (17)$$

which is in practice equivalent to the correlation function. Then equation (8) should be replaced by

the following relation

$$N(r) \propto r^{D_s}. \qquad (18)$$

Replacing the correlation function $C(r)$ with the city number $N(r)$ don't influences the spatial correlation exponent, $D_s$. It is apparent that a log-log plot of equation (18) should produce a straight line slope suggesting the fractal dimension.

According as the property of the data distribution, we choose a step length of $\triangle r$=100 kilometers. The least yardstick is 100, and the real yardstick or stride length is $r=n\triangle r$=100$n$, where $n$=1, 2, ⋯, are natural numbers. It is not difficult to calculate the number $N(r)$ with equation (17). Changing the step length $r$, we will obtain different numbers $N(r)$. Here we have a series of 40 data, and the results are list in Table 1 and are displayed in Figure 4. The double logarithmic plot shows that there exists a definite scale-free region, and the slope of the line segment in the scaling range gives the estimated value of spatial correlation dimension. Of course, the scaling range is clearly confined as the moment order is $q$=2. The higher order of correlation indicates the narrower scaling region.

Table 1 The correlation number of China's cities measured with variable yardstick

| Yardstick (r) | City Number N(r) | | | Yardstick (r) | City Number N(r) | | |
| --- | --- | --- | --- | --- | --- | --- | --- |
| | Euclidian distance | Railway distance | Highway distance | | Euclidian distance | Railway distance | Highway distance |
| 4000 | 1521 | 931 | 1028 | 2000 | 1277 | **539** | **528** |
| 3900 | 1519 | 923 | 1012 | 1900 | 1255 | **511** | **500** |
| 3800 | 1515 | 919 | 1000 | 1800 | 1187 | **453** | **474** |
| 3700 | 1511 | 909 | 988 | 1700 | **1115** | 433 | 434 |
| 3600 | 1503 | 903 | 978 | 1600 | **1061** | 415 | 408 |
| 3500 | 1499 | 895 | 962 | 1500 | **979** | 375 | 374 |
| 3400 | 1493 | 889 | 938 | 1400 | **889** | 343 | 342 |
| 3300 | 1485 | 873 | 916 | 1300 | **821** | 299 | 304 |
| 3200 | 1479 | 851 | 904 | 1200 | **741** | 263 | 274 |
| 3100 | 1473 | 841 | 882 | 1100 | **647** | 221 | 250 |
| 3000 | 1463 | 819 | **860** | 1000 | **561** | 191 | 210 |
| 2900 | 1445 | 807 | **848** | 900 | 481 | 161 | 178 |
| 2800 | 1441 | 795 | **820** | 800 | 405 | 139 | 148 |
| 2700 | 1423 | **785** | **782** | 700 | 339 | 129 | 122 |
| 2600 | 1407 | **767** | 754 | 600 | 257 | 93 | 102 |
| 2500 | 1395 | **733** | 720 | 500 | 195 | 79 | 84 |
| 2400 | 1369 | **699** | 676 | 400 | 129 | 61 | 64 |
| 2300 | 1351 | **667** | **640** | 300 | **95** | **43** | **48** |
| 2200 | 1335 | **625** | **594** | 200 | 63 | 33 | 40 |
| 2100 | 1313 | **599** | **558** | 100 | 47 | 31 | 34 |

**Note**: (1) The Euclidean distance is computed through the digital map of China by using the ESRI ArcGIS software; (2) The transportation distance data come from the *New Communications Atlas of China*. SinoMaps Press, 2003; (3) The data with bold character in this table are those falling into the scaling range.

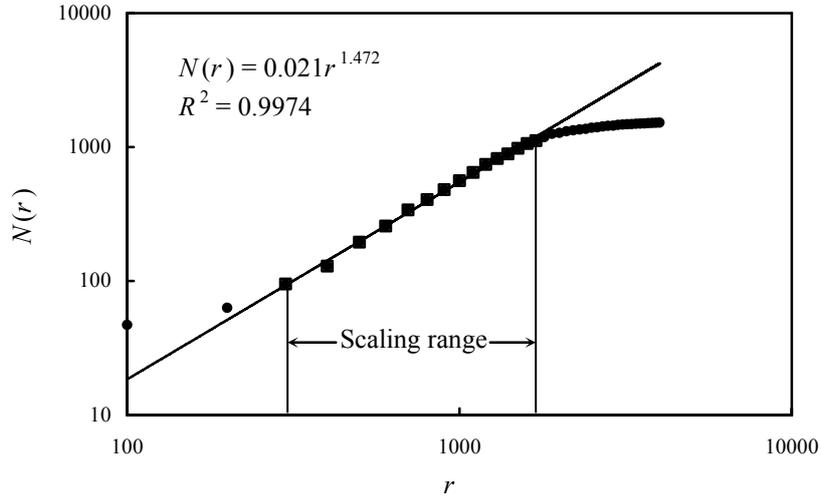

**Figure 4. The spatial correlation sum plot of Chinese cities based on Euclidean distance**

A difficult problem is to determine the limits of the scaling range objectively in the log-log plot. Two methods are employed in our study. One is the intuitionistic method, that is, one can inspect the scale-free ranges by eye before carrying out detailed calculations, and, from this visual inspection, get some general guidance to a judgment of scaling regions. However, sometimes a simple observation of the plot does not work, and then we need the second method, the regression analysis method. Firstly, we can get a straight line segment as the scale-free region using the simple intuitionistic method. Secondly, make regression analyses using the data in the line segment. If they fit well to the power function, extend the data range and test again; otherwise, reduce the range and test. Conduct the test again and again until we find the proper scaling range.

A least squares computation involving the data points falling into the scaling range yields the spatial correlation model of Chinese cities in the form

$$N(r) = 0.021 r^{1.472}.$$

The determination coefficient value is $R^2=0.9974$, and the spatial correlation dimension is estimated as $D_s=1.472$. Apparently, the scaling invariance of the spatial correlation exists only within a certain scale range of city distance. If the yardstick $r$ is so small that there is no any other city around a city, spatial correlation does not exist around the city. On the other hand, if the yardstick $r$ is so large that it is greater than the ultimate distance of urban influence, the interaction between any two cities should be ignored. This saturation effects indicates the window size. In our computation, the scaling range of cities is in between 300 *km* (lower limit)-1700 *km* (upper limit).

Similarly, we can estimate the parameter values of the spatial correlation models based on the transportation distance (Figure 5). As to the railway network, $N$=31, the number of elements of the matrix is 961. The number of data points given by equation (17) is 51, and part of the data overflowing the scaling range is omitted in Table 1. A least squares computation utilizing the data within the scale-free region yields

$$N(r) = 0.0146 r^{1.3831}.$$

The goodness of fit is $R^2$=0.9966, and the spatial correlation dimension $D_s$=1.3831. As to the highway network, $N$=34, the number of elements of the matrix is 1156. The number of data based on equation (17) is 62. (Still, Table 1 omitted part of the data overflowing the scaling region). The spatial correlation model based on the scaling range is given by a least squares analysis such as

$$N(r) = 0.024 r^{1.3155},$$

The correlation coefficient squared is $R^2$=0.9981, and the spatial correlation dimension $D_s$=1.3155. All the computation results are shown in Table 2.

**Table 2 Spatial correlation dimension, scaling range and related statistics of Chinese cities (2002)**

| Distance type | Scaling range | Total dot number | Dot number in scaling range | Correlation dimension | Goodness of fit | Spatial correlation distance |
|---|---|---|---|---|---|---|
| Euclidian Distance | 3~17 | 40 | 14 | 1.4720 | 0.9974 | 300~1700 km |
| Railway Distance | 3~27 | 51 | 24 | 1.3831 | 0.9965 | 300~2700 km |
| Highway Distance | 3~30 | 62 | 27 | 1.3155 | 0.9943 | 300~3000 km |

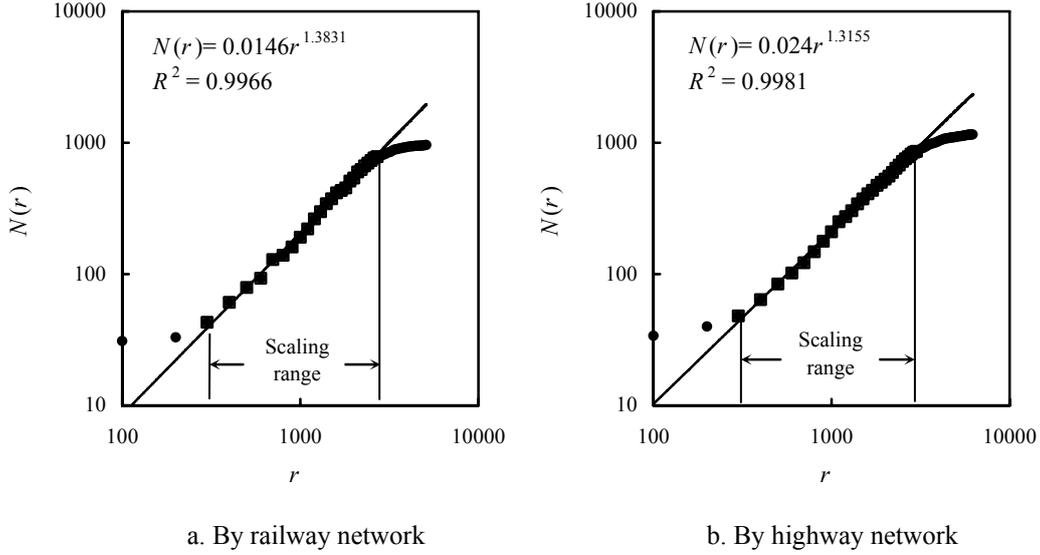

a. By railway network     b. By highway network

**Figure 5. The scaling ranges of the spatial correlation sum of Chinese cities based on transportation distance (2002)**

Now we know that the spatial correlation function of Chinese cities, measured either by the Euclidean distance or by the transportation distance, has the property of scaling invariance. This means that the spatial structure of Chinese urban system is statistically self-similar. The scaling range based on transportation distance is wider than that based on beeline distance, but the dimension value based on beeline measure is greater than that based on transportation measure. Indeed in fractal geometry it is known that when considering a tortuous and ramified network and measuring the distances $\lambda$ on the network, the number $N(\lambda)$ of points accessible is $N(\lambda) \propto \lambda^{D_c}$, where $D_c$ is the connectivity dimension (Rammal et al, 1984), which bears an analogy with the dimension related to transportation distance. Usually $D_c < D$, only for non-tortuous fractals, we have $D = D_c$. This is in accordance with the above results.

The spatial correlation dimension based on the Euclidean distance and the transportation distance can be employed to measure the development of the transportation network. For cities in a certain region, their location will not change for a short time and the Euclidean distance between any two cities will remain unchanged. However, the transportation networks usually change with the lapse of time. If the transportation between any two cities is a straight line, the transportation network of the urban system reaches a limit state. Let $D_b$ indicate the correlation dimension based on the beeline distance, and $D_t$, the dimension based on the transportation network. The more development the transportation network is, the closer $D_b$ reaches $D_t$ ($D_b \leq D_t$). With the development of the transportation, theoretically we will have

$$D_t / D_b \to 1. \tag{19}$$

Therefore, we can define an index to represent the accessibility of an urban network

$$\omega = 1 - \left|1 - \frac{D_t}{D_b}\right|. \tag{20}$$

It can be called the *development index* of a transportation network. Knowing that $0<D<2$ in theory but $1<D<2$ in practice, we have $0<\omega<1$. The closer $\omega$ reaches 1, the better connection between cities is. Based on Table 2, the development index of Chinese railway network, $\omega_r=0.9396$, and that of the highway network, $\omega_h=0.8937$. However, as we didn't conduct a computation based on an identical data set of Chinese cities, the result here is only valid to some extent.

## 4 Discussion

### 4.1 Spatial correlation of different scales

The computation in Section 3 shows that the spatial correlation function follows the power law indicative of fractal structure. It implies that the spatial distribution of cities has the property of scaling invariance, which reminds us of the possible self-organized criticality (SOC) in the urban system. Generally speaking, to a self-organizing system without evolving into critical state, the correlation function obeys the exponential law rather than the power law (Liu and Liu, 1994). The correlation function will appear in the form of power law only when the system reaches the critical state. As Hergarten (2002, P99) pointed out, "A system exhibits SOC if its phase space contains a strange attractor where events of all sizes occur, and the size distribution of these events follows a power law." This study lends further support to the notion of possible SOC in central place systems as self-organized urban networks (Chen and Zhou, 2006).

However, the scaling invariance of an urban system only exists within a certain range of scale. There is a fixed distance between any two cities. If the yardstick is too short, the probability of finding other cities around one city will be so small that there is hardly an object of spatial correlation. On the other hand, if the yardstick is too long, the corresponding distance between any pair of cities will be very far, and the correlation intensity will be too weak to be considered. A fractal city is an evolving fractal (Benguigui et al, 2000), so is a fractal urban system. The size of the scaling range suggests the development degree of the fractal structure of an urban system. Generally speaking, the more closely the evolution of an urban system approaches a self-organized critical state, the more distinct a fractal structure is. The lower limit of the scaling range indicates the shortest distance of spatial distribution of cities in an average sense, while the upper limit indicates the longest average effective distance of urban action. It can be seen from Table 2 that the average effective Euclidean distance of Chinese cities of provincial capitals is about 1700 *km*. It is about 3000 *km* measured by transportation distance. Because the railway network is likely to be made of straight lines, the effective action distance measured by railway comes between that measured by beeline and that by highway.

One question is what if we change the area coverage or the size threshold? Will the regularity in spatial correlation is still there? In order to answer this question, it is necessary to give another empirical example based on relatively small scale. We will reduce our study area to a province of China, Henan, and we will reduce the threshold of city size to the level of the main cities in this province. For simplicity, we study Henan's 17 principal cities, which construct a 17×17 matrix. The step length takes as $\triangle r$=25 km, so the smallest distance is 25, and we have $r=25n$, where $n$=1, 2, 3, …. The 17 cities are mainly connected with highway, and railway only exists between some of the cities. In this case, we conduct a computation of the Euclidean distance and the highway distance data in two different years. By means of equation (17), we get a series of yardstick and their corresponding city numbers (see Appendix 2). The double logarithmic plot of equation (18) will yield a straight line as a whole or in part, if the analytical model is correct. As expected, there is a scaling range on each log-log plot. The main modeling results are listed below

$$N(r) = 0.054 r^{1.5192}, \quad N(r) = 0.0965 r^{1.3383}, \quad N(r) = 0.0794 r^{1.3855}.$$

The analytical process shows that the spatial correlation model is also valid in a smaller region, say Henan Province of China. The main points of the analytical process can be summarized as follows. First, the scaling range measured by Euclidean distance is the smallest of the three, i.e., the result based on the transportation is better to reflect the fractal structure of an urban system. Secondly, the spatial correlation dimension based on the transport mileage increase from 1993 to 2002, accordingly, the development index of transport network rises from 0.8809 to 0.9120. The addition of traffic line to the urban network can be revealed with the index. Thirdly, for those cities with lower function and smaller size, the correlation distance indicated by the scaling range decreased a lot. The lower limit is about 50 *km* and the upper limit is about only 300 *km* (Table 3).

Table 3 Spatial correlation dimension and the scaling range of cities in Henan Province

| Distance type | Year | Scaling range | Dot number in scaling range | Correlation dimension | Goodness of fit | Correlation distance |
|---|---|---|---|---|---|---|
| Euclidean Distance | Any | 2~8 | 7 | 1.5192 | 0.9932 | 50~200 km |
| Railway Distance | 1993 | 2~14 | 13 | 1.3383 | 0.9965 | 50~350 km |
| Highway Distance | 2002 | 2~12 | 11 | 1.3855 | 0.9943 | 50~300 km |

Whether it is for a large region like China or a medium region like Henan Province, the measure of transportation distance is better than that of Euclidian distance in the case of revealing the fractal distribution of cities. As mentioned above, the transportation distance is proportional to travel time. It can be inferred that urban system in "time" space has a better fractal structure than that in conventional geographical space. Therefore, it is very necessary to develop the theory and method of map transformation using the ideas of multidimensional scaling.

### 4.2 One-point correlation and the number-radius scaling

Now let turn our attention from point-point correlation to one-point correlation of cities. This deals with the relationship between the correlation function and the regular fractal dimension. The correlation function of an urban system is related to two regular dimensions: one is the box dimension, and the other is the radial dimension (see Frankhauser, 1994). The theoretical relation between correlation dimension and box dimension is given in Section 2.3. Because the number of cities in our examples is relatively small, empirically it does not make sense to estimate the fractal dimension using the box-counting method. However, the analysis of the radius dimension can provide some useful information. In equation (5), endow *x* with a fixed value, say, let *x*=0, then the correlation function is proportional to mass function. Thus we reduce the point-point correlation function to a one-point correlation function. That is, the relationship between any two cities changes to the relationship between one city and other cities around. More straightly, in equation (17), let *j*=0, it means that we fix a city. Thus the cumulative number of correlation changes to cumulative number of cities within a radius of *r* from the given city. In this way, the dimension defined by equation (18) is not the second order correlation dimension, but the zero-order correlation dimension, i.e., the radius dimension of urban systems.

Taking city $j=0$ as the center, we can draw a series of concentric circles at regular intervals. The radius of each circle is $r$. Let $N(r)$ be the cumulative number of cities in the circles. Then we can compute the radius dimension based on the power function $N(r) \propto r^D$. The radius dimension is different from the spatial correlation dimension in that it describes the relationship between the center and its peripheral cities. There are two more simple methods to estimate the radius dimension. One is to set $r=d_{i0}$ (the subscript $j=0$ can be omitted), then we have $N(d_i) \propto d_i^D$. The other is to define an average distance

$$R_g(s) = \left(\frac{1}{s}\sum_{i=1}^{s} d_i^2\right)^{1/2}, \qquad (21)$$

where $R_g(s)$ is the *radius of gyration* (Vicsek, 1989), the serial number of cities $i=1,2, \ldots, s$; $s=1,2, \ldots, N$ ($N$ is city number). The fractal dimension based on the gyration radius can be given by

$$R_g(s) \propto s^{1/D}. \qquad (22)$$

As the distance measure is smoothed by equation (21), we can achieve a better result of dimension estimation statistically.

Take Beijing, the China's capital, as the center, then compute the radius dimension using the above two relations. The computation results from two methods share the same scaling range, i.e., $i=3\sim37$ ($N=39$). Obviously, the scaling range of the one-point correlation is much wider than that of the point-point correlation. In fact, the scaling range is narrower the higher the moment order of correlation dimension becomes. Based on the first method, we have

$$N(d_i) = 0.8607 d_i^{1.2641}.$$

The goodness of fit is $R^2=0.9932$ (Figure 6(a)). Based on the second method, we obtain

$$R_g(s) = 0.7862 s^{1/1.3104}.$$

The goodness of fit $R^2=0.9981$ (Figure 6(b)). The radius dimensions computed are almost the same, both approximate to 1.3, smaller than the former correlation dimension 1.472.

We can take each city in Figure 3 as the correlation center, and compute the corresponding one-point correlation dimension using Equations (21) and (22) based on straight line distance. It is showed that the one-point correlations of all cities except Xi'an, Hefei, Ji'nan and Qingdao have fractal characteristics. The fractal dimension ranges from 1 to 3, and mostly comes between 1 and 2, with an average of 1.6589. Several cities including Zhengzhou, Nanning, Dalian, Fuzhou and Xiamen have the structure of bifractals. Much spatio-temporal information about the evolution of urban systems can be revealed from the fractal structure and fractal dimension of the one-point correlation of each city. Such information is an important supplement to the above spatial correlation analysis. The same methods can be applied to the one-point correlation analysis based on the distance by railway and highway. As space is limited, we will not develop it in this paper. In fact, although the provincial cities in Figure 3 seem to distribute randomly, it has internal spatial order. It can be found through Cartesian transform that the distribution of these cities bears clear bilateral symmetry pattern. It is not easy to explain this kind of spatial emergence phenomenon without correlation analysis. However, the relationship between this bilateral symmetry and the dilation symmetry is a problem which needs further exploration.

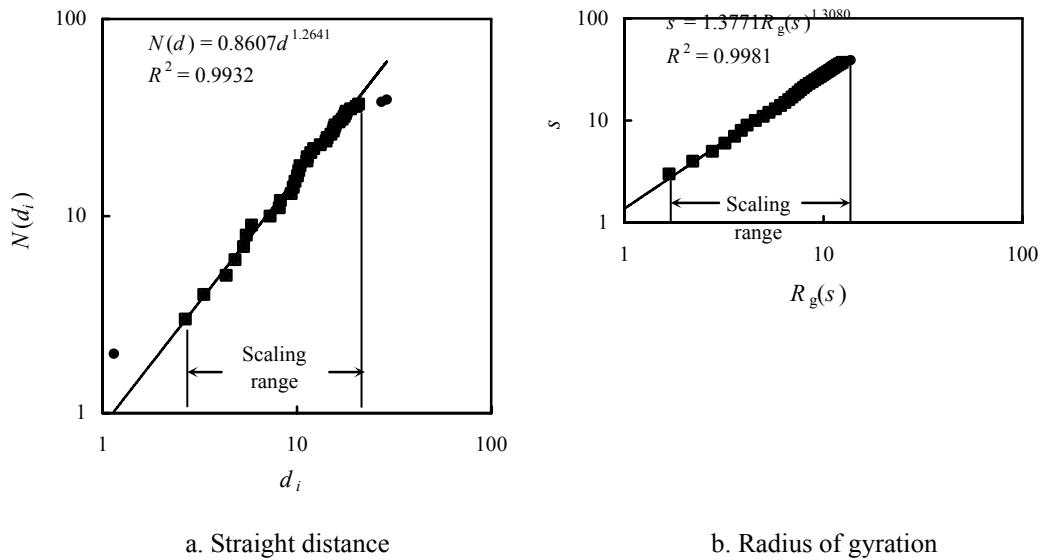

a. Straight distance       b. Radius of gyration

**Figure 6. Estimation of radius dimension based on Beijing by two methods**

**Note**: If we exchange the place of the independent variable and dependent variable, the regression result will be a little different (Chen and Zhou, 2003). Therefore, the result in Figure 6(b) differs subtly from that in text. It should be noticed that a point ($R_g(2)<1$) is out of the frame of plot and thus is not shown in Figure 6(b).

## 5 Summarization of procedure

Fractals provide new ways of looking at cities (Batty, 1995) and a new tool for the spatial analysis of cities (Frankhauser, 1998). The spatial correlation function is a useful modeling approach to researching the fractal distribution of cities, and the correlation dimension is in essence a scaling exponent of spatial autocorrelation. The main points of the analytical process in this paper can be summarized as follows.

**Firstly, determine the studied area and the threshold of the city size**. Step one is to determine the area coverage in terms of the research objects. It can be an administration district, or a region confined by a natural boundary, or even a circular or rectangular region intercepted from the map. Step two is to determine the city size threshold according to the double logarithmic plot of the city-size distribution. Whether it is a primate distribution or a rank-size distribution, the scaling range on the log-log plot has a lower limit and an upper limit, and the smallest city size should be larger than the lower limit. When all these conditions are satisfied, the threshold can be defined according as our study purpose. It can also be determined by both the city size and the importance of urban function. In order to guarantee a sample with enough number of cities, the studied area should not be too small, or the city size threshold should not be too high.

**Secondly, choose the yardstick of measurement**. The measurement can be the Euclidean distance, or the real transport mileage represented with the distance by highway or railway. The Euclidean distance is more reasonable because it is accord with the mathematical definition of the spatial correlation sum function. However, the empirical results show that the computation based on the actual transportation distance is better to describe the fractal structure of an urban system. It seems to suggest that the topological structure of an urban system in "time" space bears more practical meanings than the Euclidean geometrical structure in conventional geographic space.

**Thirdly, analyze the fractal dimension and some related parameters**. Theoretically, the spatial correlation dimension of an urban system is intervenient between 0 and 2. The larger the fractal dimension is, the more homogeneous the distribution of cities is. Otherwise, the distribution is more concentrated or nonuniform. The most valuable parameter is the development index of transportation network defined by the ratio of fractal dimensions. Given a determinate number of cities, the more closely the correlation dimension computed based on the transportation distance approaches the dimensions based on the Euclidean distance, the better accessibility among cities it suggests. With the development of transport network, the dimension based on traffic lines becomes more and more close to the dimension based on the Euclidean distance.

**Fourthly, analyze the length of the scaling range**. The spatial relations among cities are based on certain yardmeasure. If the yardstick is too small, there will be no city for correlation; otherwise, the corresponding distance will be out of the action scope of a city. Therefore, the interaction among cities is only available in certain spatial scale, which means that there exists a scaling range for effective correlation. Besides the moment order of the correlation dimension, the width of the scaling region depends on the development of the fractal urban system and the influence scope of cities. The lower limit of the scaling range reflects the character of the spatial distribution of cities while the upper limit reflects the interaction distance of cities. Analyzing the scaling range can reveal some significant information of the fractal structure of an urban system.

**Acknowledgment**: This research was sponsored by the National Natural Science Foundation of China (Grant No. 40771061). The support is gratefully acknowledged. We are grateful to Miss Jia Na for her computation of the Euclidean distance of Chinese cities.

# Appendices

## 1. Key points to understand the derivation of the correlation dimension

In the derivation process of spatial correlation dimension, the key step is the conversion from the probability based on the number of cities to the probability based on the number of small boxes. Here is a simple illustration of this process. As to Figure 2 in the text, $N$=3. The probability of finding cities in the circle around city 1 is

$$P_{11} = \frac{1}{3}, \ P_{12} = \frac{0}{3}, \ P_{13} = \frac{0}{3}.$$

Then we have a cumulative probability

$$C_1(r) = \sum_{j=1}^{3} P_{1j} = \frac{1}{3} = P_1.$$

The probability of finding cities in the circle around city 2 and city 3 is

$$P_{21}=\frac{0}{3}, \ P_{22}=\frac{1}{3}, \ P_{23}=\frac{1}{3}.$$

Thus we have

$$C_2(r)=\sum_{j=1}^{3}P_{2j}=2\times\frac{1}{3}=P_2.$$

Note that $i, j=1, 2, 3$; $k=1, 2$. As to the number of small boxes, we have $N_r=2$, $N_1(1)=1$, $N_2(2)=N_2(3)=2$. According as equation (11), we get the correlation sum as follows

$$C(r)=\frac{1}{3}\sum_{i=1}^{3}C_i(r)=\frac{1}{3}(1\times\frac{1}{3}+2\times\frac{2}{3})=(\frac{1}{3})^2+(\frac{2}{3})^2=\sum_{k=1}^{2}P_k^2.$$

This is just the result indicated by equation (12). In short, equation (11) can be transformed into equation (12) naturally. The above computation can be generalized to any number of cities.

## 2. The measurement results of spatial correlation of Henan's cities

The measurement results of spatial correlation of the main cities in Henan Province are tabulated as follows (Table A1). The bold numbers are the data falling in between the two limits of scaling range, while the italic numbers indicate the limit of spatial measurement. For Henan's 17 cities, the limit of spatial correlation number is $N(r)=17*17=289$. With the development of the network of cities from 1994 to 2003, the upper limit of correlation distance by highway becomes shorten from 575 *km* to 525 *km*.

**Table A1 The correlation number of Henan's cities measured with variable yardstick**

| Rank | Yardmeasure | Based on Euclidean distance | Based on highway distance | |
|---|---|---|---|---|
| | | 1994-2003 | 1994 | 2003 |
| 1 | 25 | 17 | 17 | 17 |
| 2 | 50 | **19** | **19** | **19** |
| 3 | 75 | **43** | 31 | 33 |
| 4 | 100 | **59** | 43 | 41 |
| 5 | 125 | **81** | 59 | 59 |
| 6 | 150 | **111** | 75 | 81 |
| 7 | 175 | **139** | 97 | 107 |
| 8 | 200 | **163** | 129 | 129 |
| 9 | 225 | 187 | **141** | **147** |
| 10 | 250 | 205 | **161** | **167** |
| 11 | 275 | 229 | **177** | **191** |
| 12 | 300 | 251 | **201** | **213** |

| 13 | 325 | 255 | **215** | 227 |
| 14 | 350 | 267 | **237** | 247 |
| 15 | 375 | 277 | 247 | 251 |
| 16 | 400 | 281 | 251 | 265 |
| 17 | 425 | 287 | 265 | 275 |
| 18 | 450 | *289* | 275 | 279 |
| 19 | 475 | *289* | 279 | 283 |
| 20 | 500 | *289* | 279 | 285 |
| 21 | 525 | *289* | 283 | *289* |
| 22 | 550 | *289* | 287 | *289* |
| 23 | 575 | *289* | *289* | *289* |